# Investigating Mathematical Models of Immuno-Interactions with Early-Stage Cancer under an Agent-Based Modelling Perspective

Grazziela P Figueredo[*1], Peer-Olaf Siebers[1] and Uwe Aickelin[1]

[1]Intelligent Modelling and Analysis Research Group, School of Computer Science, The University of Nottingham, NG8 1BB, UK

Email: Grazziela P Figueredo[*]- grazziela.figueredo@nottingham.ac.uk; Peer-Olaf Siebers - pos@cs.nott.ac.uk; Uwe Aickelin - uxa@cs.nott.ac.uk;

[*]Corresponding author

## Abstract

Many advances in research regarding immuno-interactions with cancer were developed with the help of ordinary differential equation (ODE) models. These models, however, are not effectively capable of representing problems involving individual localisation, memory and emerging properties, which are common characteristics of cells and molecules of the immune system. Agent-based modelling and simulation is an alternative paradigm to ODE models that overcomes these limitations. In this paper we investigate the potential contribution of agent-based modelling and simulation when compared to ODE modelling and simulation. We seek answers to the following questions: Is it possible to obtain an equivalent agent-based model from the ODE formulation? Do the outcomes differ? Are there any benefits of using one method compared to the other? To answer these questions, we have considered three case studies using established mathematical models of immune interactions with early-stage cancer. These case studies were re-conceptualised under an agent-based perspective and the simulation results were then compared with those from the ODE models. Our results show that it is possible to obtain equivalent agent-based models (i.e. implementing the same mechanisms); the simulation output of both types of models however might differ depending on the attributes of the system to be modelled. In some cases, additional insight from using agent-based modelling was obtained. Overall, we can confirm that agent-based modelling is a useful addition to the tool set of immunologists, as it has extra features that allow for simulations with characteristics that are closer to the biological phenomena.

**Introduction**

Advances in cancer immunology have been facilitated by the joint work of immunologists and mathematicians [1–3]. Some of the knowledge regarding interactions between the immune system and tumours is a result of using mathematical models. Most existing mathematical models in cancer immunology are based on sets of ordinary differential equations (ODEs) [2]. This approach, however, has limitations pertaining problems involving spatial interactions or emerging properties [4,5]. In addition, the analysis of ODE models is conducted at a high level of aggregation of the system entities. An alternative to ODE modelling that overcomes these limitations is systems simulation modelling. It is a set of methodologies and applications which mimic the behaviour of a real system [6–8]. Systems simulation modelling has also benefits compared to real world experimentation in immunology, including time and cost effectiveness due to the resource-intensiveness of the biological environment. Furthermore, in a simulation environment it is possible to systematically generate different scenarios and conduct experiments. In addition, the "what-if" scenarios studied in such an environment do not require ethics approval.

Agent-based modelling and simulation (ABMS) is an object-oriented system modelling and simulation approach that employs autonomous entities that interact with each other [9–11]. The focus during the modelling process is on defining templates for the individual entities (agents) being modelled and defining possible interactions between these entities. The agent behaviour is described by rules that determine how these entities learn, adapt and interact with each other. The overall system behaviour arises from the agents' individual behaviour and their interactions with other agents and the environment. For cancer immunology, it can amalgamate in vitro data on individual interactions between cells and molecules of the immune system and tumour cells to build an overview of the system as a whole [12]. Few studies, however, apply ABMS to cancer research. Although there are examples showing the success of simulation aiding advances in immunology [13–18], this set of methodologies is still not popular. There are several reasons for this: (1) ABMS is not well known in the immunology research field; (2) although simulation is acknowledged as being a useful tool by immunologists, there is no knowledge of how to use it; and (3) there is not enough trust in the results produced by simulation.

Our aim is to outline the potential contribution of ABMS to help cancer-related immune studies. In order to achieve our aim, we use a case study approach. We have chosen three well-established mathematical models describing interactions between the immune system and early-stage cancer as our test candidates. These models consist of systems of ODEs solved numerically for a time interval. By studying these models we focus on the following three research enquiries:

1. Is it possible to obtain an equivalent ABMS model based on the mathematical equations from the ODEs (i.e. can we create an object oriented model by re-using ODEs that have been created for modelling behaviour at an aggregate level)?

2. Do we get equivalent simulation outputs for both approaches?

3. What benefits could we gain by re-conceptualizing a mathematical model under an ABMS view?

**Case studies**

The case studies are chosen by considering aspects such as the population size, modelling effort, model complexity, observation of the ODEs outcome results and the number of different populations modelled. The mathematical models chosen vary largely within these aspects and therefore we can perform a more robust analysis of our experiments, as shown in Table 1.

The first case study considered is based on an ODE model involving interactions between tumour cells and generic effector cells. The second case study adds to the previous model the influence of IL-2 cytokine molecules in the immune responses of effector cells towards tumour cells. The final case study comprises an ODE model of interactions between effector cells, tumour cells, and IL-2 and TGF-$\beta$ molecules. For all case studies, the mathematical model as well as the ABMS model are presented, the outcomes are contrasted and the benefits of each approach are assessed. The models differ in terms of complexity of interactions, population sizes and the number of agents involved in the interactions (Table 1).

The remainder of this paper is organized as follows. We start with a literature review of works comparing ODEs and ABMS for different simulation domains. First, we show general work that has been carried out and then we focus on research concerned with the comparison for immunological problems. Finally, we discuss gaps in the literature regarding cancer research. In the methodology section, we introduce our agent-based modelling development process and the methods used for conducting the experimentation. In the following section, we present our case studies, comparison results and discussions. In the last section, we finally draw our overall conclusions and outline future research opportunities.

**Related Work**

In this section we describe the literature concerned with the comparison between ODE and ABMS modelling for different simulation domains. We start our review by showing general work that has been carried out to assess the differences of both approaches. Subsequently, we focus on research concerned with the comparison of the strategies for immunological problems. We found that there is a scarcity of literature comparing the two approaches for immune simulations. Furthermore, to our knowledge there is no research contrasting the approaches for the immune system and cancer interactions.

Over the past years several authors have acknowledged that little work has been done to compare both methods. In one of the pioneer studies in this area, Scholl [19] gives an overview of ODE and ABMS, describes their areas of applicability and discusses the strengths and weaknesses of each approach. The author also tries to identify areas that could benefit from the use of both methodologies in multi-paradigm simulations and concludes that there is little literature concerned with the comparison of both methodologies and their cross studies. Pourdehnad *et al.* [20] compare the two approaches conceptually by discussing the potential synergy between them to solve problems of teaching decision-making processes. The authors explore the conceptual frameworks for ODE modelling (using Systems Dynamics (SD)) and ABMS to model group learning and show the differences between the approaches in order to propose their use in a complementary way. They conclude that a lack of knowledge exists in applying multi-paradigm simulation that involves ODEs and ABMS. More recently, Stemate *et al.* [21] also compare these modelling approaches and identify a list of likely opportunities for cross-fertilization. The authors see this list as a starting point for other researchers to take such synergistic views further.

Studies on this comparison for Operations Research were also conducted. For example, Schieritz [22] and Scheritz *et al.* [23] present a cross-study of SD (which is implemented using ODEs) and ABMS. They define their features and characteristics and contrast the two methods. In addition, they suggest ideas of how to integrate both approaches. Continuing their studies, in [24] the authors then describe an approach to combine ODEs and ABMS for solving supply chain management problems. Their results show that the combined SD/ABMS model does not produce the same outcomes as ODE model alone. To understand why these differences occur, the authors propose new tests as future work.

In an application in health care, Ramandad *et al.* [25] compare the dynamics of a stochastic ABMS with those of the analogous deterministic compartment differential equation model for contagious disease spread. The authors convert the ABMS into an ODE model and examine the impact of individual heterogeneity and different network topologies. The deterministic model yields a single trajectory for each parameter set, while stochastic models yield a distribution of outcomes. Moreover, the ODE model and ABMS dynamics differ in several metrics relevant to public health. The responses of the models to policies can even differ when the base case behaviour is similar. Under some conditions, however, the differences in means are small, compared to variability caused by stochastic events, parameter uncertainty and model boundary.

An interesting philosophical analysis is conducted by Schieritz [26] analyses two arguments given in literature to explain the superiority of ABMS compared with ODEs: (1) "*the inability of ODE models to explain emergent phenomena*" and (2) "*their flaw of not considering individual diversity*". In analysing these arguments, the author considers different concepts involving simulation research in sociology. Moreover, the study identifies the theories of emergence that underlie the ODE and ABMS approaches. The author points out that "*the agent-based approach models social phenomena by modelling individuals and interactions on a lower level, which makes it implicitly taking up an individualist position of emergence; ODEs, on the other hand, without explicitly referring to the concept of emergence, has a collectivist viewpoint of emergence, as it tends to model social phenomena on an aggregate system level*". As a second part of the study, the author compares ODEs and ABMS for modelling species competing for resources to analyse the effects of evolution on population dynamics. The conclusion is that when individual diversity is considered, it limits the applicability of the ODE model. However, it is shown that "*a highly aggregate more ODE-like model of an evolutionary process displays similar results to the ABMS*". This statement suggests that there is the need to investigate further the capabilities and equivalences of each approach.

Similarly, Lorenz [27] proposes that three aspects be compared and that this helps with the choice between ODE and ABMS: structure, behaviour and emergence. Structure is related to how the model is built. The structure of a model in ODE is static, whereas in ABMS it is dynamic. In ODE, all the elements, individuals and interactions of the simulation are developed in advance. In ABMS, on the other hand, agents are created or destroyed and interactions are defined through the course of the simulation run. The second aspect (behaviour) focuses on the central generators of behaviours in the model. For ODE the behaviour generators are feedback and accumulations, while for ABMS they are micro-macro-micro feedback and interaction of the systems elements. Both methodologies incorporate feedback. ABMS, however, has feedback in more than one level of modelling. The third aspect lies in their capacity to capture emergence, which differs between the two methodologies. In disagreement with [26] mentioned earlier, the author states that ABMS is capable of capturing emergence, while the one-level structure of ODE is insufficient in that respect.

In this work we discuss the merits of ODEs and ABMS for problems involving the interactions with the immune system and early-stage cancer that can benefit from either approach. To our knowledge (and as the gap in the literature shows) such a study has not been conducted before. The differences between ODEs and ABMS when applied to classes of problems belonging to different levels of abstraction are well established in the literature [23]. However, we believe there is a range of problems that could benefit from being solved by both approaches. In addition, in many cases such as for example molecular and cellular biology, it is still not possible to use the full potential of ABMS as only the higher level of abstraction of the system is known. Another reason to investigate problems that can interchangeably benefit from both approaches is that, as many real-world scenarios, such as biological systems, constantly gather new information, the corresponding simulations have to be updated frequently to suit new requirements. For some cases, in order to suit these demands, the replacement of the current simulation approach for new developments needs to be considered. Our case study investigation seeks to provide further understanding on these problems and fill some of the gaps existing in simulations for early-stage cancer research.

**Methodology**

In this section we outline the activities and methods necessary to realise our objectives. We examine case studies of established mathematical models that describe some immune cells and molecules interacting with tumour cells. These case studies were chosen by considering aspects such as the behaviour of the entities of the model, size (and number) of populations involved and the modelling effort. The original mathematical models are built under an agent-based approach and results compared.

ABMS is capable of representing space; however, as we chose mathematical models which do not consider spatial interactions, our corresponding ABMS models do not regard space (distance) and how it would affect the simulation outcomes. The outcome samples obtained by ODEs and ABMS were statistically compared using the Wilcoxon rank-sum test to formally establish whether they are statistically different from each other. This test is applied as it is robust when the populations are not normally distributed; this is the case for the samples obtained by the ODEs and ABMS. Other approaches for assessing whether the two samples are statistically different, such as the t-test, could provide inaccurate results as they perform poorly when the samples are non-normal.

**The agent-based model development**

The agent-based models were implemented using (*AnyLogic$^{TM}$* 6.5 [28]). For the agent design we follow the steps defined in [29]: (1) identify the agents (cells and molecules), (2) define their behaviour (die, kill tumour cells, suffer apoptosis), (3) add them to an environment, and (4) establish connections to finally run the simulations, as further discussed next:

1. **Identify the possible agents**. For this purpose, we use some characteristics defined in [9]. An agent is: (1) self-contained, modular, and a uniquely identifiable individual; (2) autonomous and self-directed; (3) a construct with states that varies over time; and (4) social, having dynamic interactions with other agents that impact its behaviour. By looking at the ODE equations, therefore, the variables that are differentiated over time (their disaggregation) will either be corresponding to agents or states of one agent [29,30]. The decision whether the stock is an agent or an agent state varies depending on the problem investigated. Based on our case studies, however, we suggest that: (1) these variables preferably become states when they represent accumulations of elements from the same population; or (2) they become agents when they represent accumulations from different populations. For example, if you have an ODE $dx/dt = y$, $x$ should either be an agent or an agent state, depending on the problem context.

2. **Identify the behaviour and rules of each agent**. In our case, the agent's behaviours will be determined by mathematical equations converted into rules. Each agent has two different types of behaviours: reactive and proactive behaviours. The reactive behaviour occurs when the agents perceive the context in which they operate and react to it appropriately. The proactive behaviour describes the situations when the agent has the initiative to identify and solve an issue in the system.

3. **Implement the agents**. Based on step 2 we develop the agents. The agents are defined by using state charts diagrams from the unified modelling language (UML) [31]. With state charts it is possible to define and visualize the agents' states, transitions between the states, events that trigger transitions, timing and agent actions [4]. Moreover, at this stage, the behaviours of each agent are implemented using the simulation tool. Most of our transitions occur according to a certain rate. For our implementation, the rate is obtained from the mathematical equations.

4. **Build the simulation**. After agents are defined, their environment and behaviour previously established should be incorporated in the simulation implementation. Moreover, in this step we include parameters and events that control the agents or the overall simulation.

For example, let us consider a classical ODE which describes an early-stage tumour growth pattern [2]:

$$\frac{dT}{dt} = Tf(T) \tag{1}$$

where:

- *T* is the tumour cell population at time *t*,

- *T*(0) > 0,

- *f*(*T*) specifies the density dependence in proliferation and death of the tumour cells. The density dependence factor can be written as:

$$f(T) = p(T) - d(T) \quad (2)$$

where:

- *p*(*T*) defines tumour cells proliferation

- *d*(*T*) define tumour cells death

The expressions for *p*(*T*) and *d*(*T*) are generally defined by power laws:

$$p(T) = aT^a \quad (3)$$

$$d(T) = bT \quad (4)$$

In this case, our agent would be a tumour cell that replicates and dies according to the parameters ☐ and *3*. The tumour cell agent behaviours are "proliferate" or "die", according to the rate defined by the mathematical model. If the rate is positive, there is proliferation, otherwise, death occurs (Table 2).

The tumour cell assumes two states, alive and dead, as shown in Figure 1. In the alive state, these cells can replicate and die. If the growth rate is positive, the cell replicates according to the rate value; otherwise, it dies. There is, therefore a branch connecting the two transitions *proliferate* and *death* to the alive state. Once cells move to the final state dead, they are eliminated from the system and from the simulation.

The transition connecting the state alive to the branch is triggered by the growth rate. In the state charts, the round squares represent the states and the arrows represent the transitions between the states. Arrows within states indicate the agent actions (or behaviours) and the final state is represented by a circle.

Our agents are stochastic and assume discrete time steps to execute their actions. This, however, does not restrict the dynamics of the models, as most of our agents state transitions are executed according to certain rates - this will go in parallel with steps execution, as defined in AnyLogic [28]. The rate triggered transition is used to model a stream of independent events (Poisson stream). In case more than one transition/interaction should occur at the same time, they are executed by AnyLogic in a discrete order in the same time-step. In the next section we apply the methodology to study our case studies and compare the outcomes.

**Case 1: Interactions between Tumour Cells and Generic Effector Cells**

For the first case, a mathematical model of tumour cells growth and their interactions with general immune effector cells defined in [32] is considered. Effector cells are responsible for killing the tumour cells inside the organism. Their proliferation rate is proportional to the number of existing tumour cells. As the quantities of effector cells increase, the capacity of eliminating tumour cells is boosted. These immune cells proliferate and die per apoptosis, which is a programmed cellular death. In the model, cancer treatment is also considered. The treatment consists of injections of new effector cells into the organism. The details of the mathematical model are given in the following section.

**The mathematical model**

The interactions between tumour cells and immune effector cells can be defined by the following equations:

$$\frac{dT}{dt} = Tf(T) - d_T(T,E)$$

$$\frac{dE}{dt} = p_E(T, E) - d_E(T, E) - a_E(E) + \Phi(T) \qquad (6)$$

where

- $T$ is the number of tumour cells,
- $E$ is the number of effector cells,
- $f(T)$ is the growth of tumour cells,
- $d_T(T, E)$ is the number of tumour cells killed by effector cells,
- $p_E(T, E)$ is the proliferation of effector cells,
- $d_E(T, E)$ is the death of effector cells when fighting tumour cells,
- $a_E(E)$ is the death (apoptosis) of effector cells,

- $\Phi(T)$ is the treatment or influx of cells.

Kuznetsov model [32] defines the functions $f(T)$, $d_T(T, E)$, $p_E(E, T)$, $d_E(E, T)$, $a_E(E)$ and $\Phi(t)$ as shown below:

$$f(T) = a(1 - bT) \tag{7}$$

$$d_T(T, E) = nTE \tag{8}$$

$$p_E(E, T) = \frac{pTE}{g + T} \tag{9}$$

$$d_E(E, T) = mTE \tag{10}$$

$$a_E(E) = dE \tag{11}$$

$$\Phi(t) = s \tag{12}$$

**The agent-based model**

Two classes of agents are defined: the tumour cell and the effector cell. The agents' parameters and behaviours corresponding to each state are shown in Table 3. All behaviours are derived from the mathematical model. In the table, each agent has two different types of behaviours: reactive and proactive behaviours.

State charts are often used in ABMS to show the different states entities can be in and how they move from one state to the other (transitions). For our models we also use events, which are actions scheduled to occur in the course of the simulation. The state chart for the tumour cells is shown in Figure 2(a), in which an agent proliferates, dies with age or is killed by effector cells. In addition, at a certain rate, the tumour cells contribute to damage to effector cells. The rates defined in the transitions are the same as those from the mathematical model (Table 4). Figure 2(b) presents the effector cell agent state chart, in which either the cell is alive and able to kill tumour cells and proliferate or is dead by age or apoptosis. In the transition rate calculations, the variable *TotalTumourCells* corresponds to the total number of tumour cell agents; and the variable *TotalEffectorCells* is the total number of effector cell agents. In the simulation model, apart from the agents, there is also an event – namely, treatment – which produces new effector cells with a rate defined by the parameter *s*.

**Experimental design for the simulations**

Four scenarios were investigated. The scenarios have different death rates of tumour cells (defined by parameter *b*), different effector cells apoptosis rates (defined by parameter *d*) and different treatments (parameter *s*). The values for these parameters were obtained from [2] (see Table 5). In the first three scenarios, cancer treatment was considered, while the fourth case did not consider any treatment. The simulation for the ABMS was run fifty times and the mean values are displayed as results. The mathematical model outcomes display values for a time period of one hundred days and therefore the same interval was determined for the ABMS.

**Results and discussion**

In the first scenario results, shown in Figure 3, the behaviour of the tumour cells appears similar for both ODEs and ABMS. However, the Wilcoxon test rejected the similarity hypothesis for both outcomes, as shown in Table 6. The reason for this test pointing out that the outcomes differ is that tumour cells for the ODE model decreased asymptotically towards to zero, while the ABMS behaviour is discrete and therefore **reached zero**. Furthermore, the variances observed in the ABMS curve, given its stochastic characteristic, also influenced the Wilcoxon test results. The number of effector cells for both simulations follow the same pattern, although the numbers are not the same due to the agents variability. This variability is very evident with regards to the effector cells population for two main reasons: (1) for this first case study the size of the populations involved is relatively small, which increases the impacts of stochasticity in the outcomes; and (2) the ODE system changes the amount of cells overtime in a continuous fashion, which means that in this simulation fractions of cells are considered. ABMS does not consider fraction of cells - a cell either is alive or dead. This is implemented as a boolean indicator and corresponds to the real world, where fractions of cells could obviously not exist. Considering the above explanations we conclude that for this scenario the ABMS outcomes seem more realistic, as in biological experiments cells are also atomic entities and stochastic variability occurs.

The results for the second scenario seem similar for effector cells, as shown in Figure 4, which was confirmed by the Wilcoxon test (Table 6). The results for the tumour cells are visibly not the same. Regarding the ODE simulation, in the first ten days the tumour cells population first decreases and then grows up to a value of 240 cells, in which the growth reaches a steady-state. The initial decrease of tumour cells is also observed in the ABMS outcomes. After ten days, however, there is a smaller cellular increase and a steady-state is not observed. Similar to the previous scenario, the simulation curve presents an erratic behaviour throughout the simulation days. There is, however an unexpected decay of tumour cells over time. This is explained by the individual characteristics of the agents and their growth/death rates attributed to their instantiation. As the death rates of tumour cells agents are defined according to the mathematical model, when the tumour cell population grows, the newborn tumour cells have higher death probabilities, which leads to a considerable number of cells dying out. This indicates that the individual behaviour of cells can lead to a more chaotic behaviour when compared to the aggregate view observed in the ODE simulation.

For scenarios 3 and 4, shown in Figures 5 and 6 respectively, the results for both approaches differ completely. Moreover, with regard to the tumour cells curve, the differences are even more evident. The ODEs outcomes for scenario 3 reveal that tumour cells decreased as effector cells increased, following a predator-prey trend curve. For the ABMS, however, the number of effector cells decreased until a value close to zero was reached, while the tumour cells numbers varied differently from the ODEs results. As we discussed for the previous scenarios, the predator prey-pattern observed in the ODE simulation was only possible due to its continuous character. In the ODE simulation outcome curve for the effector cells it is possible to observe, for instance, that after sixty days the number of effector cells ranges between one and two. These values could not be reflected in the ABMS simulation and therefore the differences occur.

In scenario 4, although effector cells appear to decay in a similar trend for both approaches, the results for tumour cells vary widely. In the ODE simulation, the numbers of effector cells reached a value close to zero after twenty days and then increased to a value smaller than one. For the ABMS simulation, however, these cells reached zero and never increased again.

Similar to scenarios 2 and 3, the continuous ODE simulation outcomes contrasted with discrete agents caused the different outcomes. Furthermore, as occurred in scenario 2, the individual behaviour and rates attributed to the cells seemed to have an impact in the growth of tumours.

**Summary**
An ODE model of tumour cells growth and their interactions with general immune effector cells was considered for re-conceptualization using ABMS. Four scenarios considering small population numbers were investigated and, for only one of them, the ABMS results were similar to the mathematical model. The differences observed were explained by the way each simulation approach is implemented, which includes their data representation and processing. ODE simulations deal with continuous values for the entities whereas ABMS represents discrete agents. Furthermore, the stochastic behaviour of the ABMS and how it affects small populations is not present in the ODEs. It also appears that the individual interactions between populations in the ABMS leads to a more chaotic behaviour, which does not occur at a higher aggregate level. The result analysis also reveals that conceptualizing the ABMS model from the mathematical equations does not always produce the same outcomes. One alternative to obtain better matching results would be the development of an agent-based model, which is not based on the rates defined in the ODE model, but using real data (available or collectable) or some form of parameter calibration.

**Case 2: Interactions Between Tumour Cells, Effector Cells and Cytokines IL-2**

The second case study investigated is concerned with a mathematical model for the interactions between tumour cells, effector cells and the cytokine IL-2. This is an extension of the previous study, since it considers IL-2 as molecules that will mediate the immune response towards tumour cells. They will interfere on the proliferation of effector cells according to the number of tumour cells in the system. Treatment is now applied in two different ways, by injecting effector cells or injecting cytokines.

**The mathematical model**

The mathematical model used in case 2 is obtained from [33]. The model's equations illustrate the non-spatial dynamics between effector cells (E), tumour cells (T) and the cytokine IL-2 ($I_L$), described by the following differential equations:

$$\frac{dE}{dt} = cT - \mu E + \frac{p_1 E I_L}{g_1 + I_L} + s_1$$

Equation 13 describes the rate of change for the effector cell population E [33]. Effector cells grow based on recruitment ($cT$) and proliferation ($\frac{p_1 E I_L}{g_1 + I_L}$). The parameter $c$ represents the antigenicity of the tumour cells (T) [33,34]. $\mu$ is the death rate of the effector cells. $p_1$ and $g_1$ are parameters used to calibrate the recruitment of effector cells and $s_1$ is the treatment that will boost the number of effector cells.

$$\frac{dT}{dt} = a(1 - bT) - \frac{a_a ET}{g + T}$$

Equation 14 describes the changes that occur in the tumour cell population T over time. The term $a(1 - bT)$ represents the logistic growth of T ($a$ and $b$ are parameters that define how the tumour cells will grow) and $\frac{a_a ET}{g_2+T}$ is the number of tumour cells killed by effector cells. $a_a$ and $g_2$ are parameters to adjust the model.

$$\frac{dI_L}{dt} = \frac{p_2 ET}{g_3+T} - \mu_3 I_L + s_2 \qquad (15)$$

The IL-2 population dynamics is described by Equation 15. $\frac{p_2 ET}{g_3+T}$ determines IL-2 production using parameters $p_2$ and $g_3$. $\mu_3$ is the IL-2 loss. $s_2$ also represents treatment. The treatment is the injection of IL-2 in the system.

**The agent-based model**

The populations of agents are therefore the effector cells, tumour cells and IL-2 and their behaviour is shown in Table 7. The state charts for each agent type are shown in Figure 7. The ABMS model rates corresponding to the flow values in the ODEs model are given in Table 8. In the transition rate calculations, the variable *TotalTumour* corresponds to the total number of tumour cell agents, the variable *TotalEffector* is the total number of effector cell agents and *TotalIL* 2 is the total number of IL-2 agents.

In the simulation model, apart from the agents, there are also two events:

1. *TreatmentS*1, which adds effector cell agents according to the parameter $s1$

2. *TreatmentS*2, which adds IL-2 agents according to the parameter $s2$

**Experimental design for the simulation**

The experiment was conducted assuming the same parameters as those of the mathematical model (Table 9). For the ABMS model, the simulation was run fifty times and the average outcome value for these runs was collected. Each run simulated a period equivalent to six hundred days, following the same time span used for the numerical simulation of the mathematical model.

**Results and discussion**

The results obtained are shown in Figures 8, 9 and 10 for effector cells, tumour cells and IL-2 respectively. The ABMS was validated by comparing its outputs with those produced by the ODEs. As the figures reveal, the results for all populations are very similar; the growth and decrease of all populations occur at similar times for both approaches. Furthermore because of the large population sizes (around $10^4$), ABMS model curves have minor erratic behaviour, which corroborates to the similar patterns observed in the outcomes.

These similarities are also confirmed by the Wilcoxon test results presented in Table 10. The table shows the p-values obtained with a 5% significance level. For the effector and tumour cells, the p-value was higher than 0.5, which indicates that the test failed to reject the null hypothesis that the outcomes were similar.

**Summary**

A mathematical model for the interactions between tumour cells, effector cells and the cytokine IL-2 was considered to investigate the potential contribution of building the model under an ABMS perspective. Experimentation shows that results are very similar, which is explained by the large population sizes considered in the experiments. In further experiments, the same model was also run under small population sizes and the results for the simulations were different due to stochasticity and the approaches particularities, as discussed in the previous case study. Regarding the use of computational resources for larger data sets, ABMS was far more time- and memory-consuming than the ODEs.

## Case 3: Interactions Between Tumour Cells, Effector Cells, IL-2 and TGF-$\beta$

The third case study is based on the mathematical model of Arciero *et al.* [34], which consists of a system of ODEs describing interactions between tumour cells and immune effector cells, as well as the immune-stimulatory and suppressive cytokines IL-2 and TGF-$\beta$. According to Arciero *et al.* [34] TGF-$\beta$ stimulates tumour growth and suppresses the immune system by inhibiting the activation of effector cells and reducing tumour antigen expression. The mathematical model, together with further details on the interactions studied is introduced in the following section.

**The mathematical model**

The mathematical model we use in case 2 is obtained from [33]. The model's equations illustrate the non-spatial dynamics between effector cells (E), tumour cells (T), IL-2 (*I*) and TGF-$\beta$ (S) cytokines. The model is described by the following differential equations:

$$\frac{dE}{dt} = \frac{cT}{1+\gamma S}(p_{1EI}) \left( \mu_1 - \underline{\phantom{x}} q_1 S \right) \quad (16)$$

Equation 16 describes the rate of change for the effector cell population E. According to [34], *effector cells are assumed to be recruited to a tumour site as a direct result of the presence of tumour cells*. The parameter $c$ in $\frac{cT}{1+\gamma S}$ represents the antigenicity of the tumour, which measures the ability of the immune system to recognize tumour cells. The presence of TGF-$\beta$ (*S*) reduces antigen expression, thereby limiting the level of recruitment, measured by inhibitory parameter $\alpha$. The term $\mu_1 E$ represents loss of effector cells due to cell death, and the proliferation term $\left(\frac{p_1 E I}{g_1 + I}\right)$ presence of the cytokine IL-2 and is decreased when the cytokine TGF-$\beta$ is present. $p_1$ is the maximum rate of effector cell proliferation in the absence of TGF-$\beta$, $g_1$ and $q_2$ are half-saturation constants, and $q_1$ is the maximum rate of anti-proliferative effect of TGF-$\beta$ $\left(\frac{q_2 \Sigma}{\Sigma}\right)$ asserts that effector cell proliferation depends on the

$$\frac{dT}{dt} = aT\left(1 - \frac{T}{K}\right) - \frac{a_\alpha ET}{g_2 + T} + \frac{p_2 ST}{g_3 + S} \tag{17}$$

Equation 17 describes the dynamics of the tumour cell population. The term $aT\left(1 - \frac{T}{K}\right)$ represents logistic growth dynamics with intrinsic growth rate $a$ and carrying capacity $K$ in the absence of effector cells and TGF-$\beta$. The term $\frac{a_\alpha ET}{\gamma_2 + T}$ is the number of tumour cells killed by effector cells. The parameter $a_\alpha$ measures the strength of the immune response to tumour cells. The third term $\frac{\pi_2 \Sigma T}{\gamma_3 + \Sigma}$ accounts for the increased growth of tumour cells in the presence of TGF-$\beta$. $p_2$ is the maximum rate of increased proliferation and $g_3$ is the half-saturation constant, which indicates a limited response of tumour cells to this growth-stimulatory cytokine [34].

$$\frac{dI}{dt} = \frac{p_3 ET}{(g_4 + T)(1 + \alpha S)} - \mu_2 I \tag{18}$$

The kinetics of IL-2 are described in equation 18. The first term $\frac{\pi_3 ET}{(\gamma_4 + T)(1+\alpha\Sigma)}$ represents IL-2 production which reaches a maximal rate of $p_3$ in the presence of effector cells stimulated by their interaction with the tumour cells. In the absence of TGF-$\beta$, this is a self-limiting process with half-saturation constant $g_4$ [34]. The presence of TGF-$\beta$ inhibits IL-2 production, where the parameter $\alpha$ is a measure of inhibition. Finally, $\mu_2 I$ represents the loss of IL-2.

$$\frac{dS}{dt} = \frac{p_4 T^2}{\alpha_2 + T^2} - \mu_3 S \tag{19}$$

Equation 19 describes the rate of change of the suppressor cytokine, TGF-$\beta$. According to [34], *experimental evidence suggests that TGF-$\beta$ is produced in very small amounts when tumours are small enough to receive ample nutrient from the surrounding tissue. However, as the tumour population grows sufficiently large, tumour cells suffer from a lack of oxygen and begin to produce TGF-$\beta$ in order to stimulate angiogen-esis and to evade the immune response once tumour growth resumes. This switch in TGF-$\beta$* production is modelled by term $\frac{\pi_4 T^2}{\sigma_2 + T^2}$, where $p_4$ is the maximum rate of TGF-$\beta$ production and $\sigma$ is the critical tumour cell population in which the switch occurs. The decay rate of TGF-$\beta$ is represented by the term $\mu_3 S$.

### The agent-based model

Our agents are the effector cells, tumour cells, IL-2 and TGF-$\beta$ and the behaviour of each agent is shown in Table 11. The state charts for each agent type were developed, as illustrated in Figure 11. The ABMS model rates corresponding to the mathematical model are given in Table 12. In the transition rate calculations, the variable *TotalTumour* corresponds to the total number of tumour cell agents; the variable *TotalEffector* is the total number of effector cell agents, *TotalIL*2 is the total number of IL-2 agents and *TotalTGFBeta* is the total TGF-$\beta$ agents. This model does not include any events.

### Experimental design for the simulation

The experiment was conducted assuming the same parameters as those defined for the mathematical model (Table 13). For the ABMS model, the simulation was run fifty times and the average outcome value for these runs was collected. Each run simulated a period equivalent to six hundred days, following the time interval used for the numerical simulation of the mathematical model.

### Results and discussion

Results for case 3 ODEs and ABMS simulations are provided in Figures 12, 13, 14 and 15. Outcomes demonstrate that the behaviour of the curves for effector cells, tumour cells and IL-2 in both paradigms is similar, although the starting time for the growth of populations for the ABMS varies for each run. In the figures corresponding to the ABMS results, therefore, ten distinct runs were plotted to illustrate the variations.

For most ABMS runs the pattern of behaviour of the agents is the same as that obtained by the ODEs. For a few runs, however, the populations decreased to zero, indicating that it is not always possible to obtain similar results with both approaches.

The differences observed occur for two reasons: (1) the ABMS stochasticity and (2) the agents individual behaviour and their interactions. While ODEs always use the same values for the parameters over the entire population aggregate, ABMS rates vary with time. Each agent therefore is likely to have distinct numbers for their probabilities. The agents individual interactions, which give raise to the overall behaviour of the system, are also influenced by the scenario determined by the random numbers used. By running the ABMS multiple times with different sets of random numbers, the outcomes vary according to these sets. For the ODEs, on the contrary, multiple runs always produce the same outcome, as random numbers are not considered.

In addition, the unexpected patterns of behaviour found the the ABMS results are the consequence of the agents individual interactions and their chaotic character. We believe that these unexpected patterns obtained with ABMS should be further investigated by specialists to determine if they are realistic and plausible to happen in biological experiments.

Regarding the TGF-$\beta$ outcomes, the ODEs results reveal numbers smaller than one, which is not possible to achieve with the ABMS. The results for the simulations regarding these molecules are therefore completely different and the ABMS results are always zero.

Figures 16, 17, 18 and 19 contrast the ODEs results with the closest results obtained from ABMS. For all experiments, ABMS demanded far more computational resources than the ODEs simulation runs.

**Summary**

The third case study comprised interactions between effector cells, tumour cells and two types of cytokines, namely IL-2 and TGF-$\beta$. There were two important aspects observed in the ABMS outcomes. The first observation is that the TGF-$\beta$ population was not present in the simulation when using the mathematical model's parameters, as its numbers are real values smaller than one. This indicates that there is the need of further model validation with real data in order to check which paradigm outcome is closer to reality. The second aspect observed is that ABMS produces extra patterns of population behaviour (extreme cases) distinct from that obtained by the mathematical model. This could in turn lead to the discovery of other real-world patterns, which would otherwise not be revealed by deterministic models.

**Conclusions**

In the literature there is little work related to the application of ABMS to cancer research. ODE models are more frequently used instead. Immune research could benefit from ABMS as an alternative to ODE modelling that overcomes some of its limitations regarding emergence, individual memory, adaptation and spatial localization. In this paper, our aim was to outline the possible contribution of ABMS to early-stage cancer research by immunologists. In order to achieve our aim, three research questions were defined: (1) Is it possible to obtain an equivalent agent-based model from the original mathematical formulation? (2) Do the outcomes differ? (3) Are there any benefits of using one method compared to the other? To answer these questions we chose three well established mathematical models. These differ in terms of population sizes, types of agents involved and nature of the interactions. A summary of our case studies and findings is depicted in Table 14.

Case study 1 was concerned with the use of ODEs to model interactions with general immune effector cells and tumour cells. The objective of this model is to observe these two populations evolving overtime and to evaluate the impacts of cancer treatment in their dynamics. Four different scenarios regarding distinct sets of parameters were investigated and in the first three scenarios treatment was included. The ABMS produced very different results for most scenarios. The outcomes from ODEs and ABMS only resembled for Scenario 1. It appears that two major characteristics of this model influenced the differences obtained: (1) The small quantities of individuals considered in the simulations (especially regarding the effector population size, which was always smaller than ten) that significantly increase the variability of the ABMS; and (2) The original mathematical model considers fractional population sizes (smaller than one) which is impossible to be considered in ABMS. In addition to this particular model's characteristics, for any mathematical model considering cyclic intervals of growth or decay of populations observed in our studies, the corresponding curves in the ABMS outcomes are more accentuated, given the fact that ODEs changes quantities continuously whereas ABMS varies discretely. Small numbers do not allow to recreate predator-prey patterns in stochastic models, as such models need to be perfectly balanced to work. Stochasticity in small models does not allow such balance and therefore might produce chaos.

Case study 2 referred to the investigation of the interactions between effector cells, cytokines IL-2 and tumour cells, and only one scenario was considered. ODEs and ABMS simulations also produced very similar results. As populations' sizes had a magnitude of 104 individuals, the ABMS erratic behaviour in the outcomes was not evident, which contributed to the outcomes similarity. The differences observed in the curves were explained by the continuous numbers produced by the ODEs versus discrete values from ABMS.

Case study 3 added complexity to the previous case study by establishing a mathematical model including the influence of the cytokine TGF-$\beta$ in the interactions between effector cells, cytokines IL-2 and tumour cells. The simulation outcomes for the ABMS were mostly following the same pattern as that produced by the ODEs; however there were some alternative outcomes where the patterns of behaviour demonstrated a total extermination of tumour cells by the first two hundred days. This indicates that for this case study the ABMS results are more informative, as they illustrate another set of possible dynamics that should be validated through further immune experimentation.

In response to our research questions, we conclude that not everything modelled in ODEs can be implemented in ABMS (e.g. no half agents); however – this does not matter if population sizes in the original model definition are large enough. In addition, population size has a positive impact on result similarity. The bigger the population, the closer the simulation outputs. Finally, ABMS can contribute additional insight, as due to its stochastic nature it can produce different results (normal and extreme cases). Further, variability in the graph is closer to the real world, although knowing the underlying pattern might be more useful. ODEs therefore also have an advantage as they show more clearly underlying patterns in the output (as for example predator-prey pattern) .

In the future, we want to investigate new case studies and systematically determine when phenomena such as agent-based stochasticity mostly influences on the outcome differences and in which circumstances extreme cases occur. In addition, with regard to extreme cases, it is necessary to gain additional insights of (1) how frequent these extreme cases occur and (2) wether there is any relation between the frequency of occurrence of these cases in the simulation and in the real-world. For example, we could count the appearance of these unusual cases (as a measure of system stability or robustness of the solution) when running the experiments 10,000 times. This could help immunologists defining vaccination strategies and appropriateness of cancer treatments by making them aware of the possible outcome scenarios and how frequently they occur.

## List of Abbreviations

**ODE** Ordinary differential equation

**ABMS** Agent-based modelling and simulation

## Authors Contributions

Grazziela Figueredo developed the agent-based simulation models, carried out the experiments and drafted the manuscript. Peer-Olaf Siebers helped drafting the manuscript and reviewed it critically for important intellectual content. Uwe Aickelin reviewed the manuscript and gave the final approval of the version to be published.

# Figures
**Figure 1 - Tumour cell agent**

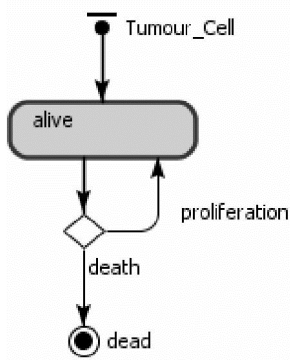

**Figure 2 - ABMS state charts for case 1**

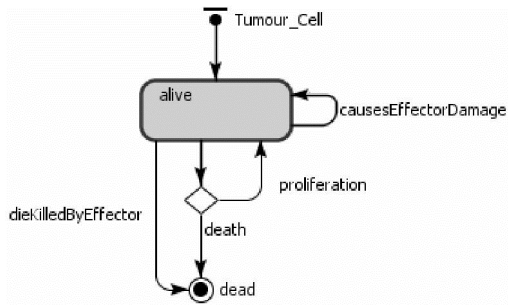 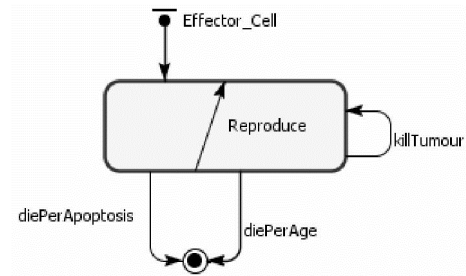

    (a) Tumour cell agent         (b) Effector cell agent

**Figure 3 - Results for case 1 scenario 1**

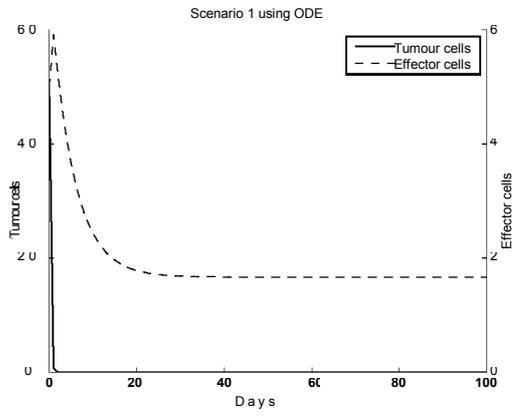
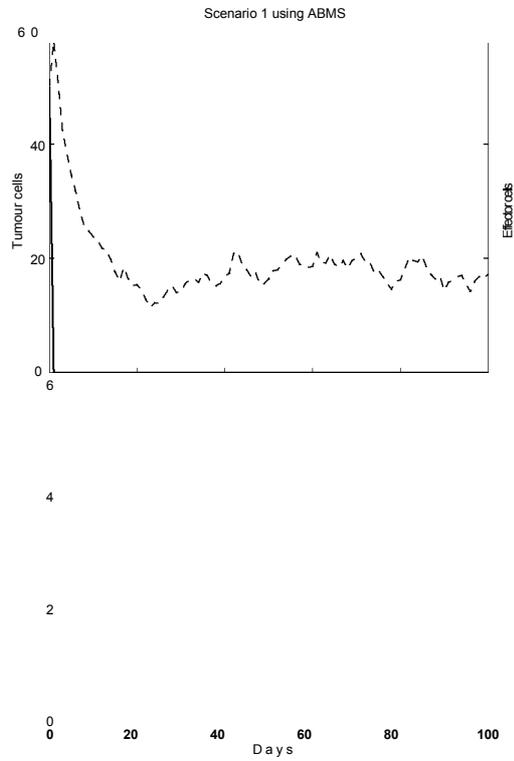

**Figure 4 - Results for case 1 scenario 2**

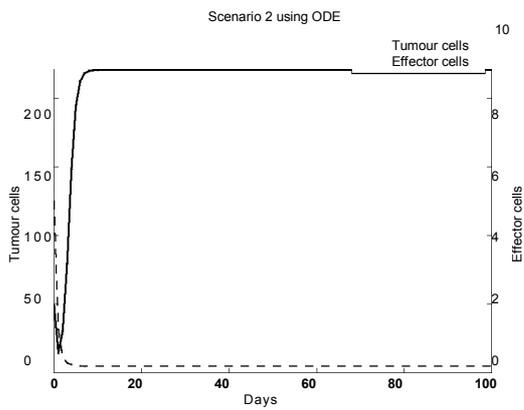
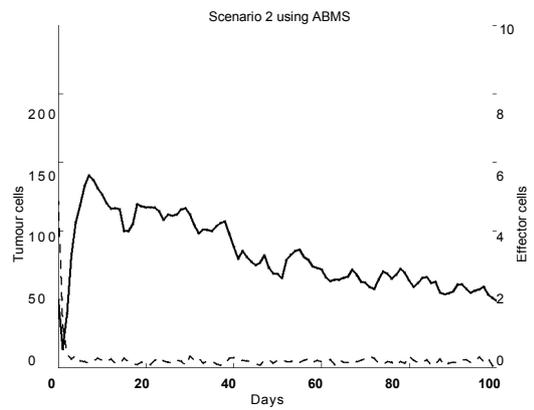

**Figure 5 - Results for case 1 scenario 3**

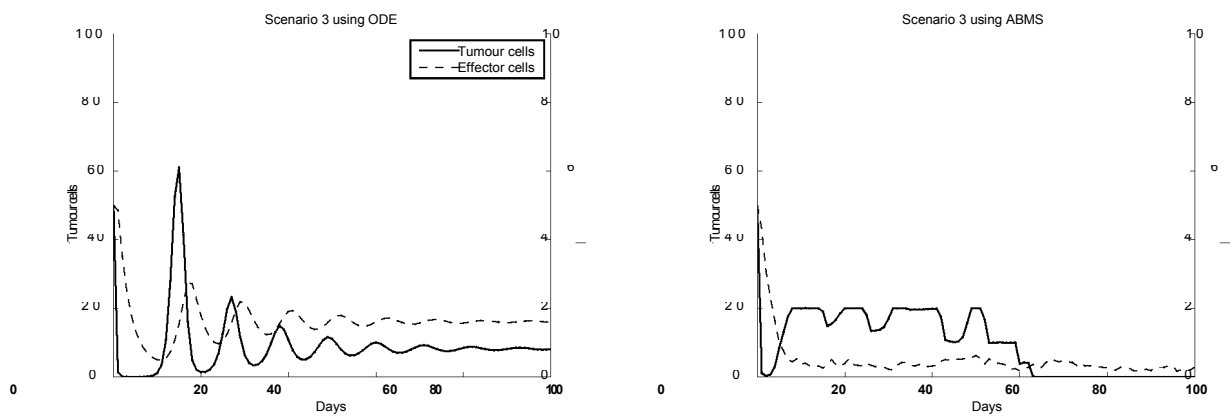

**Figure 6 - Results for case 1 scenario 4**

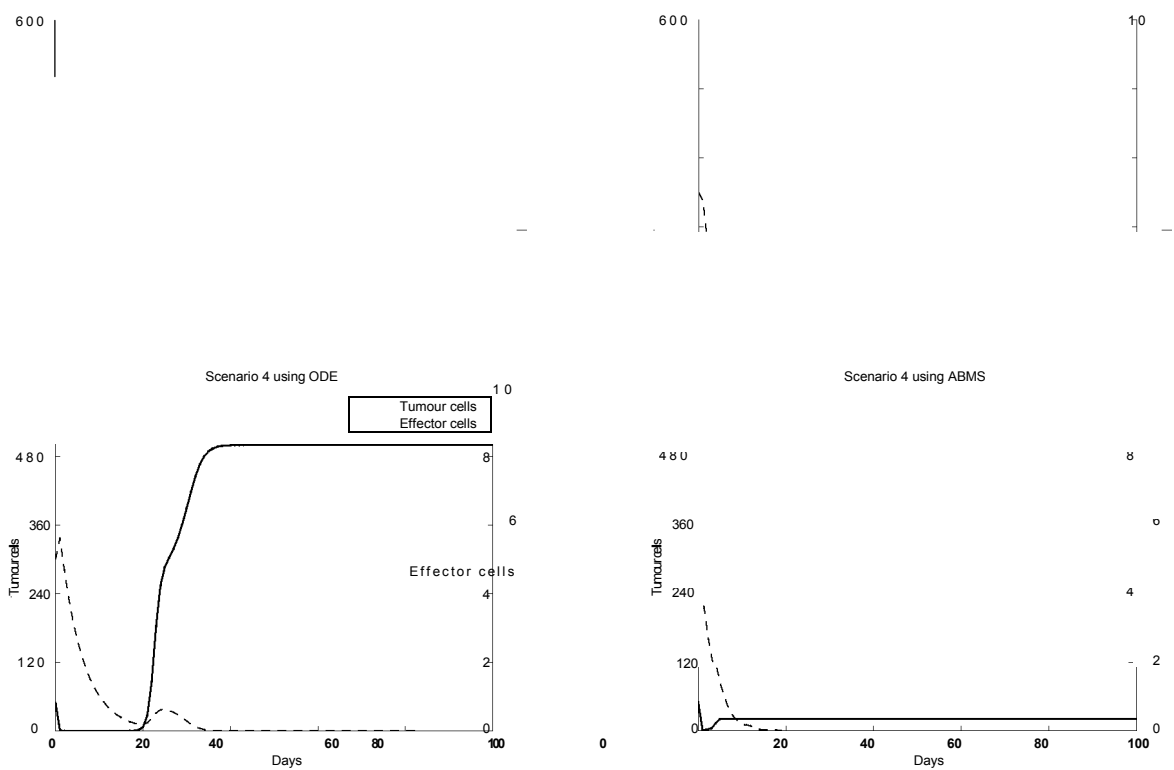

Figure 7 - **ABMS state charts for the agents of case 2**

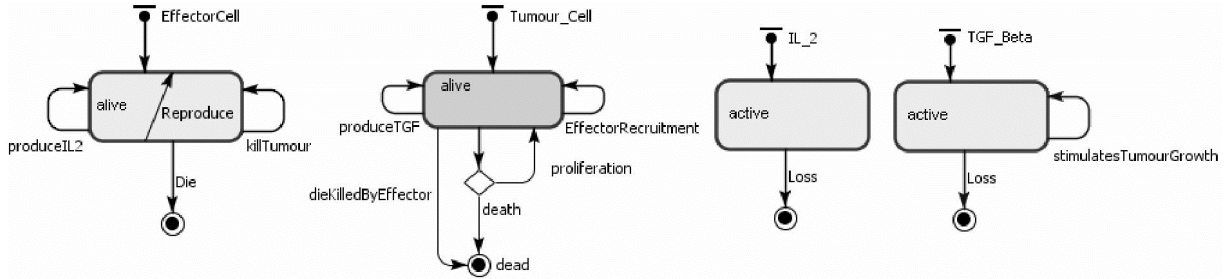

**Figure 8 - ODEs and ABMS results for effector cells of case 2**

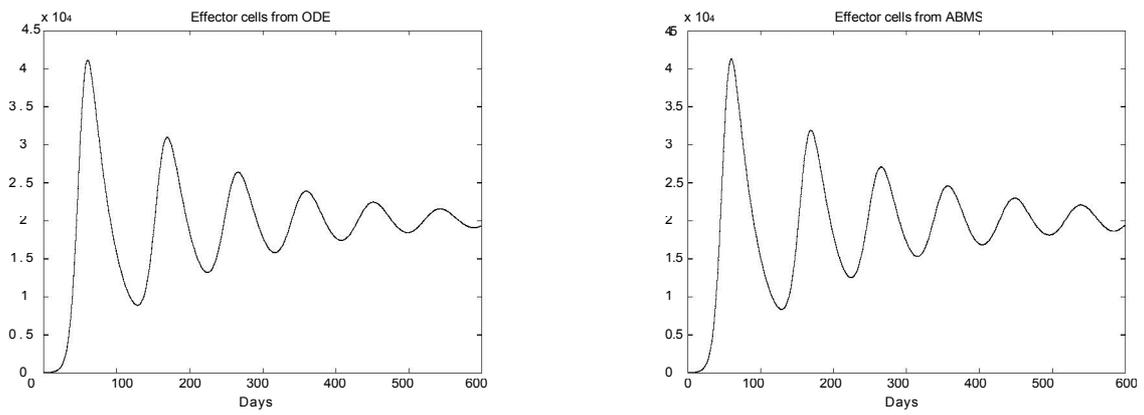

**Figure 9 - ODEs and ABMS results for tumour cells of case 2**

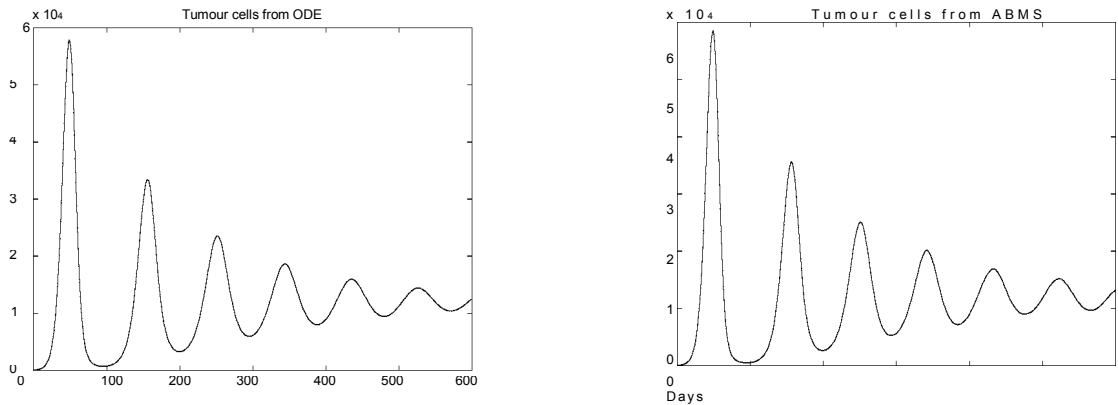

**Figure 10 - ODEs and ABMS results for IL2 of case 2**

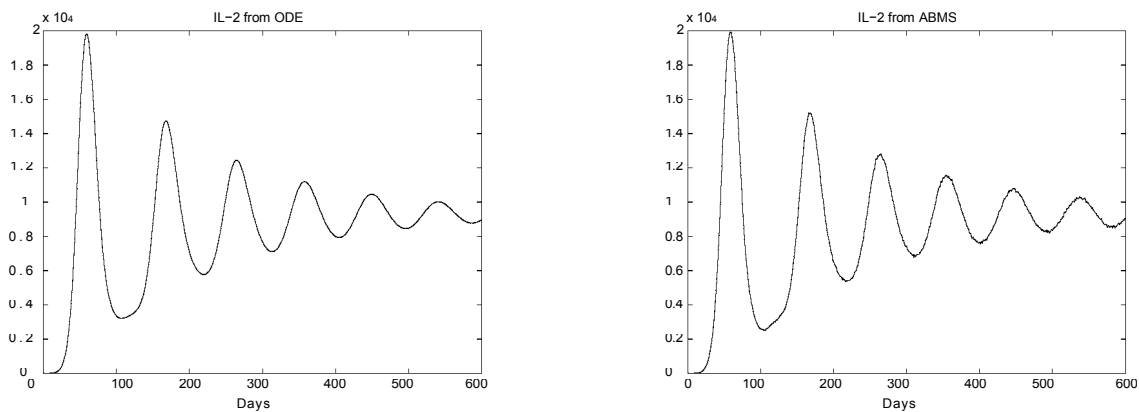

**Figure 11 - ABS state charts for the agents of case 3**

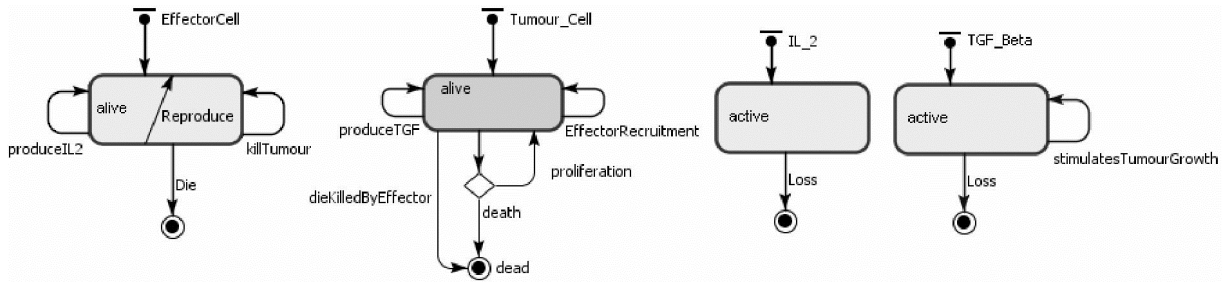

**Figure 12 - ODEs and ten runs of ABMS results for effector cells of case 3**

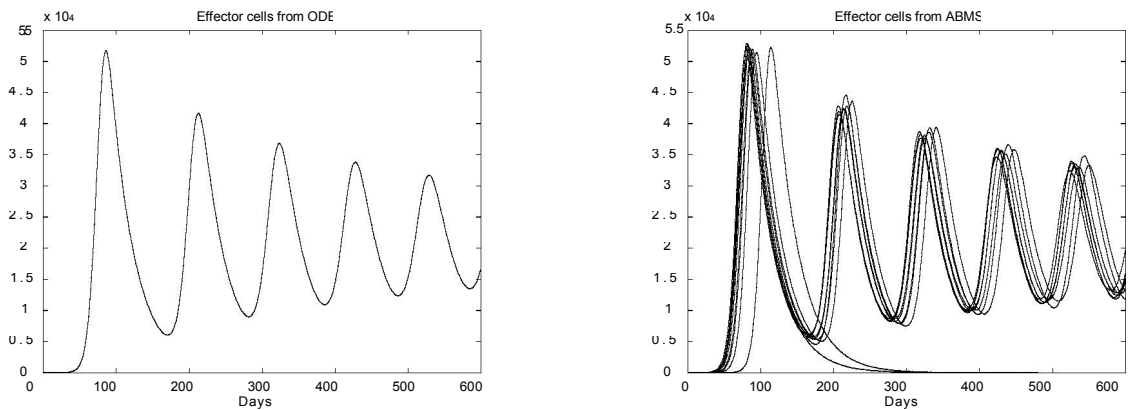

**Figure 13 - ODEs and ten runs of ABMS results for tumour cells of case 3**

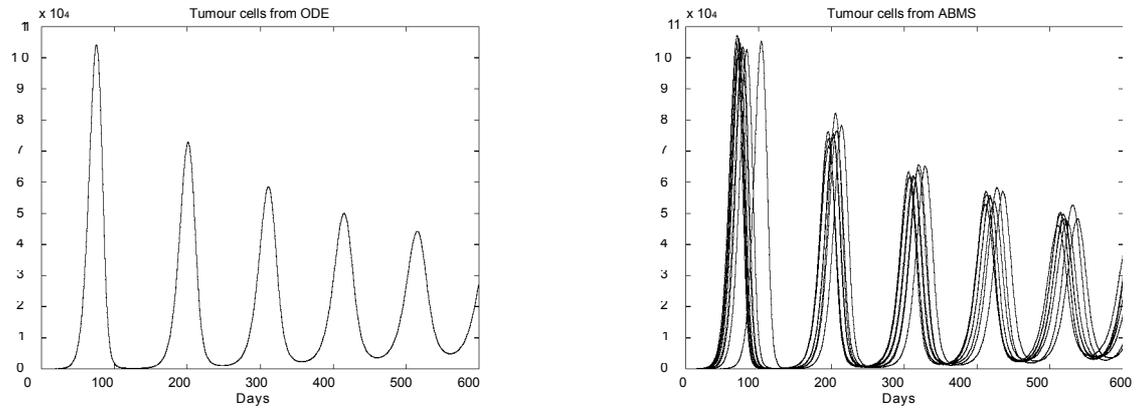

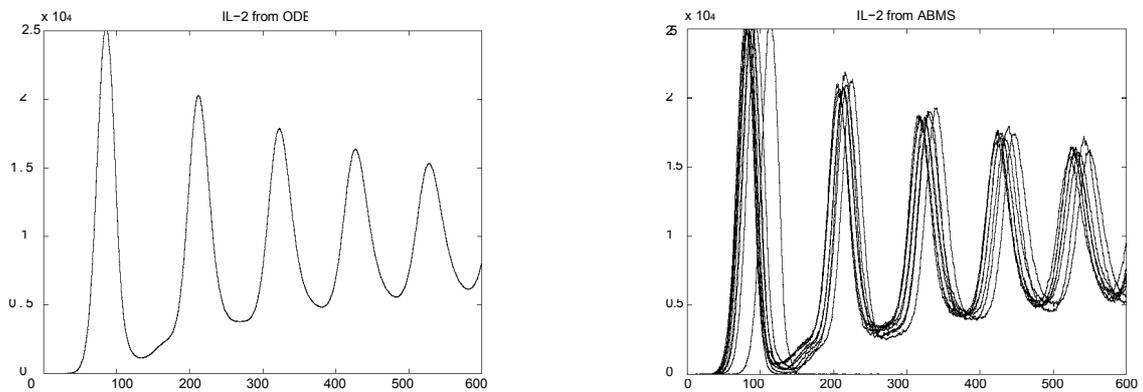

**Figure 14 - ODEs and ten runs of ABMS results for IL-2 of case 3**

**Figure 15 - ODEs and ten runs of ABMS results for TGF-$\beta$ of case 3**

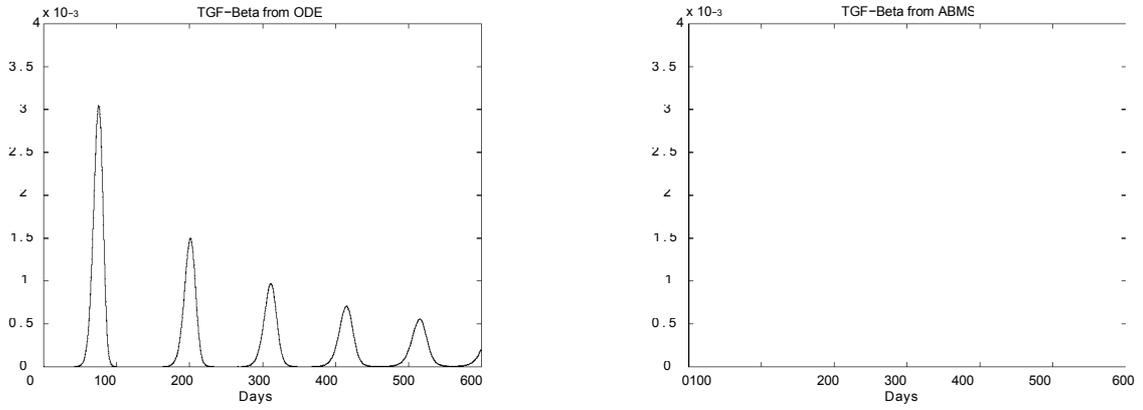

**Figure 16 - ODEs and ABMS results for effector cells of case 3**

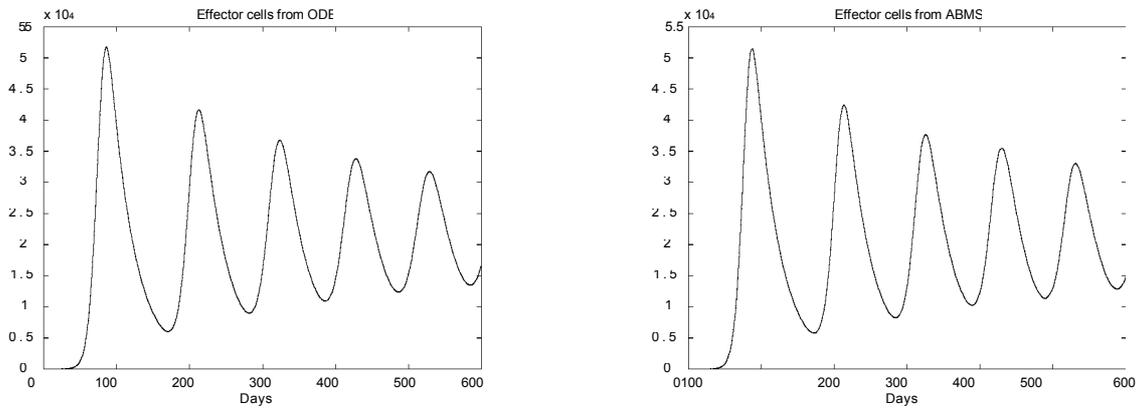

**Figure 17 - ODEs and ABMS results for tumour cells of case 3**

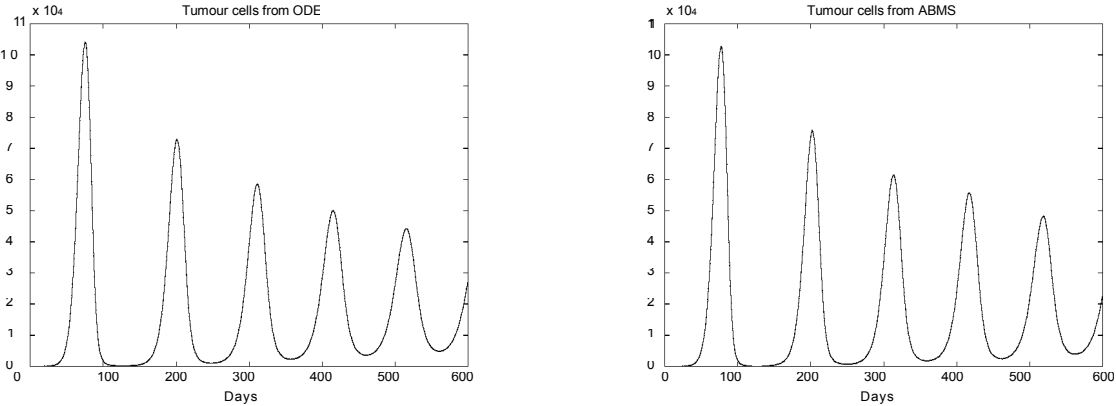

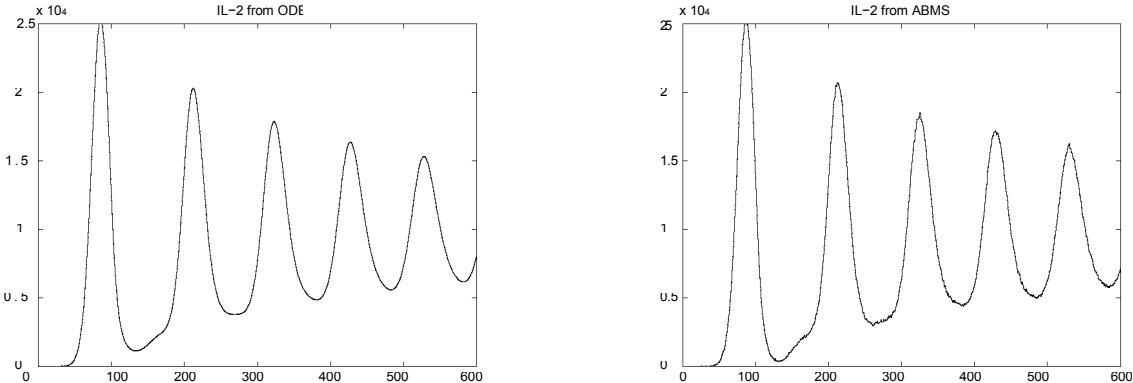

**Figure 18 - ODEs and ABMS results for IL-2 of case 3**

**Figure 19 - ODEs and ABMS results for TGF-$\beta$ of case 3**

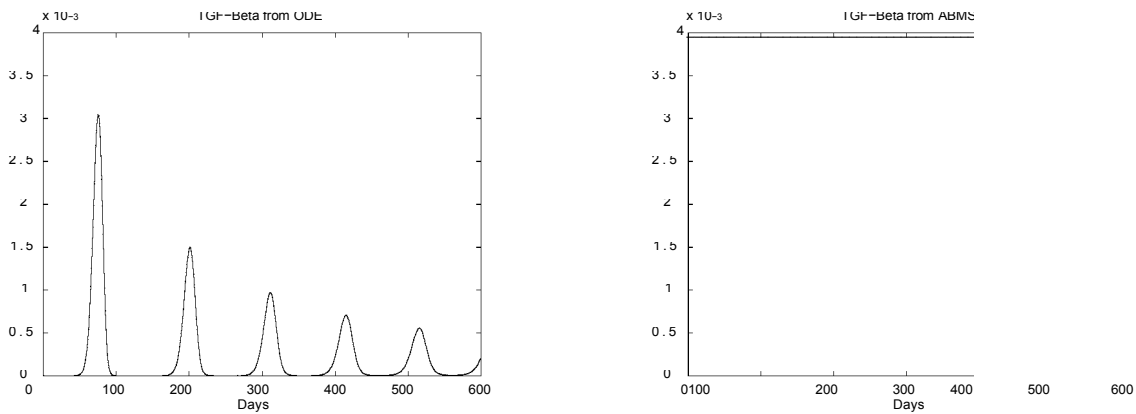

## Tables
**Table 1 - Case studies considered**

| Case Study | Number of populations | Population size | Complexity |
|---|---|---|---|
| 1) Tumour/Effector | 2 | 6 to 600 | Low |
| 2) Tumour/Efector/IL-2 | 3 | $10^4$ | Medium |
| 3) Tumour/Effector/IL-2/ TGF-$\beta$ | 4 | $10^4$ | High |

**Table 2 - Agents' parameters and behaviours for the tumour growth model**

| Parameters | Reactive behaviour | Proactive behaviour |
|---|---|---|
| *a*, *alpha*, *b* and *beta* | Dies if *rate* < 0 | Proliferates if *rate* > 0 |

**Table 3 - Agents' parameters and behaviours for case 1**

| Agent | Parameters | Reactive behaviour | Proactive behaviour |
|---|---|---|---|
| Tumour Cell *m* | *a* and *b* | Dies (with age) | |
| | *a* and *b* | | Proliferates |
| | | | Damages effector cells |
| | *n* | Dies killed by effector cells | |
| Effector Cell *p* | *m* | Dies (with age) | |
| | *d* | Dies per apoptosis | |
| | and *g* | | Proliferates |
| | *s* | Is injected as treatment | |

**Table 4 - Transition rates calculations from the mathematical equations for case 1**

| Agent | Transition | Mathematical equation | Transition rate |
|---|---|---|---|
| Tumour Cell | proliferation | $aT(1 - Tb)$ | $a - (TotalTumour.b)$ |
| | death | $aT(1 - Tb)$ | $a - (TotalTumour.b)$ |
| | dieKilledByEffectorCells | $nTE$ | $n.TotalEffectorCells$ |
| | causeEffectorDamage | $mTE$ | $m$ |
| Effector Cell | Proliferation | $\frac{pTE}{g+T}$ | $\frac{p.TotalTumourCells}{g+TotalTumourCells}$ |
| | DieWithAge | $dE$ | $d$ |
| | DiePerApoptosis | $mTE$ | message from tumour |

**Table 5 - Simulation parameters for different scenarios of case 1. For the other parameters, the values are the same in all experiments, i.e.** $a = 1.636$, $g = 20.19$, $m = 0.00311$, $n = 1$ **and** $p = 1.131$.

| Scenario | b     | d      | s      |
|----------|-------|--------|--------|
| 1        | 0.002 | 0.1908 | 0.318  |
| 2        | 0.004 | 2      | 0.318  |
| 3        | 0.002 | 0.3743 | 0.1181 |
| 4        | 0.002 | 0.3743 | 0      |

**Table 6 – Wilcoxon test with** 5% **significance level comparing case 1 simulation results**

| Implementation | Cells    | Scenario (p-value) | | | |
|----------------|----------|--------|--------|--------|--------|
|                |          | 1      | 2      | 3      | 4      |
| *ABS*          | *Tumour*   | 0      | 0      | 0.8591 | 0      |
|                | *Effector* | 0.3789 | 0.6475 | 0      | 0      |
| *ABS - Fix 1*  | *Tumour*   | 0      | 0      | 0      | 0.0011 |
|                | *Effector* | 0      | 0.3023 | 0      | 0      |
| *ABS - Fix 2*  | *Tumour*   | 0      | 0      | 0      | 0      |
|                | *Effector* | 0      | 0      | 0      | 0      |

**Table 7 – Agents' parameters and behaviours for case 2**

| Agent | Parameters | Reactive behaviour | Proactive behaviour |
|-------|------------|--------------------|--------------------|
| Effector Cells1 | $mu2$ | Dies | |
| | $p1$ and $g1$ | | Reproduces |
| | $c$ | Is recruited | |
| | | Is injected as treatment | |
| | $p2$ and $g3$ | | Produces IL-2 |
| | $aa$ and $g2$ | | Kills tumour cells |
| Tumour Cell | $a$ and $b$ | Dies | |
| | $a$ and $b$ | | Proliferates |
| | $aa$ and $g2$ | Dies killed by effector cells | |
| | $c$ | | Induces effector recruitment |
| IL-2 | $p2$ and $g3$ | Is produced | |
| | $mu3$ | Is lost | |
| | $s2$ | Is injected | |

**Table 8 - Transition rates calculations from the mathematical equations for case 2**

| Agent | Transition | Mathematical equation | Transition rate |
|---|---|---|---|
| Effector Cell | Reproduce | $\frac{p_1 I_L E}{g_1 + IL_2}$ | $\frac{p_1 \cdot TotalIL_2 \cdot TotalEffector}{g_1 + TotalIL_2}$ |
| | Die | $\mu_2 E$ | $mu2$ |
| | killTumour | $\frac{a_a ET}{g_2 + T}$ | $aa \frac{TotalTumour}{g_2 + TotalTumour}$ |
| | ProduceIL2 | $\frac{p_2 ET}{g_3 + T}$ | $\frac{p_2 \cdot TotalTumour}{g_3 + TotalTumour}$ |
| Tumour Cell | Reproduce | $aT(1 - bT)$ | $a - (TotalTumour \cdot b)$ |
| | Die | $aT(1 - bT)$ | $a - (TotalTumour \cdot b)$ |
| | DieKilledByEffector | $\frac{a_a TE}{g_2 + T}$ | message from effector |
| IL-2 | Loss | $\mu_3 I_L$ | $mu3$ |

**Table 9 - Parameter values for case 2**

| Parameter | Value |
|---|---|
| a | 0.18 |
| b | 0.000000001 |
| c | 0.05 |
| aa | 1 |
| g2 | 100000 |
| s1 | 0 |
| s2 | 0 |
| mu2 | 0.03 |
| p1 | 0.1245 |
| g1 | 20000000 |
| p2 | 5 |
| g3 | 1000 |
| mu3 | 10 |

**Table 10 - Wilcoxon test with 5% significance level comparing the results from the ODEs and ABMS for case 2**

| Population | p |
|---|---|
| Effector | 0.7231 |
| Tumour | 0.5710 |
| IL2 | 0.4711 |

**Table 11 - Agents' parameters and behaviours for case 3**

| Agent | Parameters | Reactive behaviour | Proactive behaviour |
|---|---|---|---|
| Effector Cell | $mu1$ | Dies | |
| | $p1$, $g1$, $q1$ and $q2$ | | Reproduces |
| | $c$ | Is recruited | |
| | $aa$ and $g2$ | | Kills tumour cells |
| Tumour Cell | $a$ | Dies | |
| | $a$ | | Proliferates |
| | $aa$ and $g2$ | Dies killed by effector cells | |
| | $g3$ and $p2$ | Has growth stimulated | |
| | $p4$ and $tetha$ | | Produces TGF-$3$ |
| | $c$ | | Induces effector recruitment |
| IL-2 | $alpha$, $p3$ and $g4$ | Is produced | |
| | $mu2$ | Is lost | |
| TGF-$3$ | $p4$ and $tetha$ | Is produced | |
| | $mu3$ | Is lost | |
| | $p2$ and $g3$ | | Stimulates tumour growth |

**Table 12 - Transition rates calculations from the mathematical equations for case 3**

| Agent | Transition | Mathematical equation | Transition rate |
|---|---|---|---|
| Effector Cell | Reproduce | $p_1 E \left(1 - \dfrac{\pi_1 E}{g_1 + S}\right)$ | $p_1 \text{Total}E \left(1 - \dfrac{\pi_1 \times \text{TotalIL2}}{g_1 + \text{TotalTGF}}\right)$ |
|  | Die | $\mu_1 E$ | $mu1$ |
|  | ProduceIL2 | $\dfrac{\pi_3 TE}{(\gamma_4 + T)(1 + \alpha S)}$ | $\dfrac{\pi_3 \cdot \text{TotalTumour}}{(\gamma_4 + \text{TotalTumour})(1 + \alpha \cdot \text{TotalTGF})}$ |
|  | KillTumour | $\dfrac{\alpha'' TE}{\gamma_2 + T}$ | $\dfrac{\alpha_\alpha \times \text{TotalTumour} \times \text{TotalEffector}}{\gamma_2 + \text{TotalTumour}}$ |
| Tumour Cell | Reproduce | $\left(aT \left(1 - \dfrac{T}{1000000000}\right)\right)$ | $\left(\text{TotalTumour} \cdot a \left(1 - \dfrac{\text{TotalTumour}}{1000000000}\right)\right)$ |
|  | Die | $\left(aT \left(1 - \dfrac{T}{1000000000}\right)\right)$ | $\left(\text{TotalTumour} \cdot a \left(1 - \dfrac{\text{TotalTumour}}{1000000000}\right)\right)$ |
|  | DieKilledByEffector | $\dfrac{\alpha'' TE}{\gamma_2 + T}$ | message from effector |
|  | ProduceTGF | $\dfrac{\pi_4 T^2}{theta^2 + T^2}$ | $\dfrac{\pi_4 \cdot \text{TumourCells}}{theta^2 + \text{TumourCells}^2}$ |
|  | EffectorRecruitment | $\dfrac{\chi T}{1 + S}$ | $\dfrac{\chi}{1 + \gamma \cdot \text{TotalTGF}}$ |
| IL-2 | Loss | $\mu_2 I$ | $mu2$ |
| TGF-β | Loss | $\mu_3 S$ | $mu3$ |
|  | stimulates TumourGrowth | $\dfrac{\pi_2 T}{\gamma_3 + S}$ | $\dfrac{\pi_2 \cdot \text{TotalTGF}}{\gamma_3 + \text{TotalTGF}}$ |

**Table 13 - Parameter values for case 3**

| Parameter | Value |
|---|---|
| a | 0.18 |
| aa | 1 |
| alpha | 0.001 |
| c | 0.035 |
| g1 | 20000000 |
| g2 | 100000 |
| g3 | 20000000 |
| g4 | 1000 |
| gamma | 10 |
| mu1 | 0.03 |
| mu2 | 10 |
| mu3 | 10 |
| p1 | 0.1245 |
| p2 | 0.27 |
| p3 | 5 |
| p4 | 2.84 |
| q1 | 10 |
| q2 | 0.1121 |
| theta | 1000000 |

**Table 14 - Summary of findings**

| Case Study | Outcome of the comparison | Explanation | Population size |
|---|---|---|---|
| 1) Tumour/Effector | • Most results were different | • It appears that variabilities in small populations have major impacts in the outcomes | Varied from 6 to 600 |
| 2) Tumour/Efector | • Results were statistically /IL-2 the same | • Large populations<br>• Less variability in the agents' populations | $10^4$ |
| 3) Tumour/Effector/ IL-2/ TGF-$9$ | • Different runs with outcome variations<br>• Simulations produced alternative scenarios<br>• The behaviour of the curves is less erratic for agents | • Agent-based stochasticity<br>• New scenarios need further investigations to assess their feasibility<br>• Large numbers of agents | $10^4$ |